\documentclass[twocolumn,showkeys,aps,prb,showpacs]{revtex4-1}
\usepackage{graphicx}
\usepackage[CJKbookmarks,dvipdfm,colorlinks,linkcolor=black,citecolor=black]{hyperref}

\begin{document}

\title{ Predicted septuple-atomic-layer Janus  $\mathrm{MSiGeN_4}$ (M=Mo and W) monolayers with Rashba spin splitting and high electron carrier
mobilities}

\author{San-Dong Guo$^{1,2}$, Wen-Qi Mu$^{1}$, Yu-Tong Zhu$^{1}$, Ru-Yue Han$^{1}$ and Wen-Cai Ren$^{3,4}$}
\affiliation{$^1$School of Electronic Engineering, Xi'an University of Posts and Telecommunications, Xi'an 710121, China}
\affiliation{$^2$Key Laboratary of Advanced Semiconductor Devices and Materials, Xi'an University of Posts and Telecommunications, Xi'an 710121, China }
\affiliation{$^3$Shenyang National Laboratory for Materials Science, Institute of Metal Research,
Chinese Academy of Science, 110016 Shenyang, Liaoning, P. R. China}
\affiliation{$^4$School of Materials Science and Engineering, University of Science and Technology of China,
Shenyang 110016, P. R. China}
\begin{abstract}
 Janus two-dimensional (2D) materials have attracted much attention due to possessing  unique properties  caused by their out-of-plane
asymmetry, which have been achieved in many 2D families.  In this work,   the Janus monolayers are predicted in  new 2D $\mathrm{MA_2Z_4}$ family by means of first-principles calculations, $\mathrm{MoSi_2N_4}$ and $\mathrm{WSi_2N_4}$ of which  have been  synthesized  in experiment(\textcolor[rgb]{0.00,0.00,1.00}{Science 369, 670-674 (2020)}). The predicted  $\mathrm{MSiGeN_4}$ (M=Mo and W) monolayers exhibit  dynamic,  thermodynamical and mechanical stability, and they are  indirect band-gap semiconductors.
The inclusion of spin-orbit coupling (SOC) gives rise to the  Rashba-type spin splitting,  which is observed in the valence bands, being different from common conduction bands.  Calculated results show valley polarization at the edge of the conduction bands  due to SOC together with inversion symmetry breaking.   It is found that $\mathrm{MSiGeN_4}$ (M=Mo and W) monolayers have high electron mobilities.
Both  in-plane and much weak out-of-plane piezoelectric polarizations can be observed,  when  a uniaxial strain in the basal plane is applied.
The values of piezoelectric strain  coefficient
$d_{11}$ of the Janus $\mathrm{MSiGeN_4}$ (M=Mo and W) monolayers fall between those of the
$\mathrm{MSi_2N_4}$ (M=Mo and W) and $\mathrm{MGe_2N_4}$ (M=Mo and W) monolayers, as expected. It is proved that strain can tune the positions of valence band maximum (VBM) and conduction band minimum (CBM), and enhance  the the strength of conduction bands convergence caused by compressive strain.  It is also found that tensile biaxial  strain can enhance
 $d_{11}$ of  $\mathrm{MSiGeN_4}$ (M=Mo and W) monolayers, and the compressive strain can improve the  $d_{31}$ (absolute values).
Our predicted  $\mathrm{MSiGeN_4}$ (M=Mo and W) monolayers as derivatives of 2D $\mathrm{MA_2Z_4}$ family   enrich  Janus   2D materials,  and can motivate related experimental works.

\end{abstract}
\keywords{Janus  monolayers, Carrier mobility, Piezoelectronics}

\pacs{71.20.-b, 77.65.-j, 72.15.Jf, 78.67.-n ~~~~~~~~~~~~~~~~~~~~~~~~~~~~~~~~~~~Email:sandongyuwang@163.com}

\maketitle

\section{Introduction}
The exploration of graphene\cite{q6} enormously  promotes the search for new 2D materials  both in experiment and in theory, which have potential  applications in the field of optoelectronics,  spintronics,  valleytronics and energy conversion and storage.
Numerous 2D materials have been found, including transition metal
chalcogenides (TMDs), group-VA, group IV-VI,  group-IV, transition metal
carbides/nitrides (Mxenes), $\mathrm{Cr_2Ge_2Te_6}$, $\mathrm{Mn_2C_6Se_{12}}$ and $\mathrm{Mn_2C_6S_6Se_6}$   monolayers\cite{q7,q8,q9,p1,q6-1,q6-1-1,q6-1-2,q6-1-3,q6-1-4,q6-2,q10,q11}. The unique crystal structure  together with strong SOC in monolayer
TMDs demonstrates coupled spin-valley physics\cite{q6-5}, and the buckled
honeycomb structure plus  strong SOC  can give rise to quantum spin Hall
(QSH) and quantum anomalous Hall (QAH) effects in
a particular type of 2D Xene\cite{q6-5-1,q6-5-2}.  An emerging class of 2D materials (Janus 2D materials) have  currently attracted increasing
attention due to unique crystal structures, which lack the reflection symmetry with respect to the
central  atomic  layer\cite{q6-5-3}.  In these 2D Janus materials, the strong Rashba spin splitting,
second harmonic generation response and out-of-plane  piezoelectric polarizations can be  achieved\cite{q6-5-3}.
Many Janus 2D materials have been proposed, such as Janus graphene, asymmetrically functionalizing silicene monolayer, Janus TMDs, Janus transition-metal oxides,  PtSSe, TiXY (X/Y=S, Se and Te), VSSe, SnSSe and  Janus group-III monochalcogenide $\mathrm{M_2XY}$ (M=Ga, In; X/Y=S, Se, Te)\cite{zs-1,zs-2,zs-3,zs-4,zs-5,zs-6,zs-7,zs-8,zs-9}. Recently, Janus monolayer MoSSe has been successfully achieved  by different
experimental strategies\cite{p1,p1-new} with additional out-of-plane piezoelectric
coefficient\cite{p2,p2-new}.

\begin{figure*}
  \includegraphics[width=15.0cm]{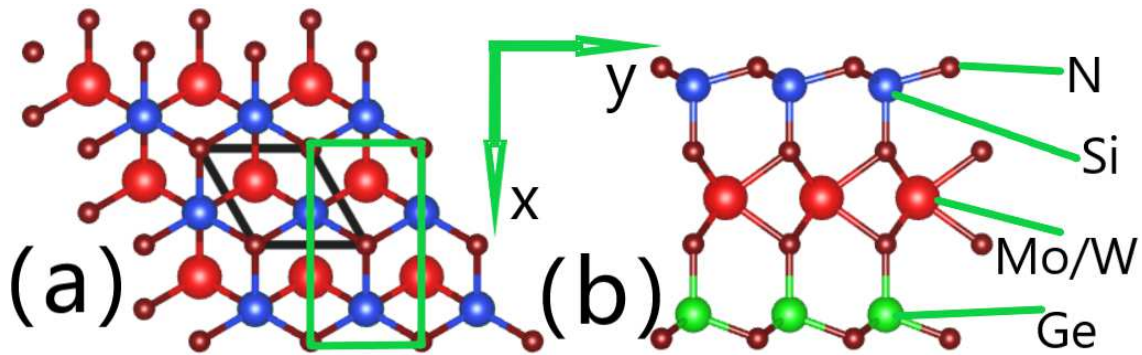}
  \caption{(Color online) The top view (a) and side view (b) crystal structure of  $\mathrm{MSiGeN_4}$ (M=Mo and W) monolayer. The rhombus primitive cell  and the rectangle supercell are marked by  black and green lines.}\label{t0}
\end{figure*}

\begin{figure}
  \includegraphics[width=8cm]{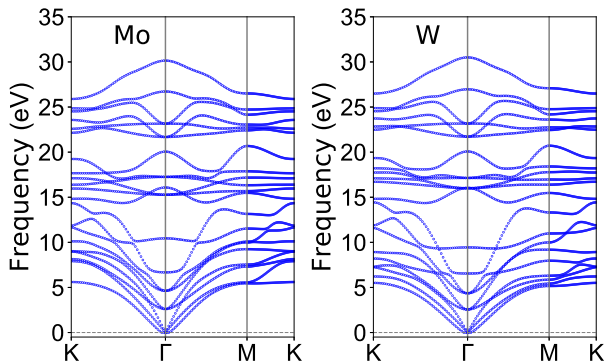}
\caption{(Color online) The phonon band dispersions  of   $\mathrm{MSiGeN_4}$ (M=Mo and W). }\label{ph}
\end{figure}

Recently, by chemical vapor deposition (CVD), the septuple-atomic-layer
2D $\mathrm{MoSi_2N_4}$ and $\mathrm{WSi_2N_4}$ have been  synthesized\cite{msn}, which opens up a new 2D material family.
The density functional theory (DFT) calculations predict many similar 2D materials  with a general formula of $\mathrm{MA_2Z_4}$, where M represents an early transition metal
(W, V, Nb, Ta, Ti, Zr, Hf, or Cr), A is Si or Ge, and Z stands for N, P, or As\cite{msn}. In quick succession, by intercalating $\mathrm{MoS_2}$-type  $\mathrm{MZ_2}$ monolayer into InSe-type  $\mathrm{A_2Z_2}$ monolayer,  twelve kinds of 2D family $\mathrm{MA_2Z_4}$ are proposed with $\alpha_i$ and  $\beta_i$ ($i$=1 to 6) phases with diverse
properties from semiconductor to topological insulator to Ising superconductor\cite{m20}. Intrinsic piezoelectricity in monolayer  $\mathrm{MSi_2N_4}$ (M=Mo, W, Cr, Ti, Zr and Hf) has been predicted by the first-principle calculations\cite{m21}.  It is also predicted that the strain can effectively tune the electronic properties of  $\mathrm{VSi_2P_4}$ monolayer, and it undergoes  ferromagnetic metal (FMM) to spin-gapless  semiconductor (SGS) to ferromagnetic semiconductor (FMS) to SGS to ferromagnetic half-metal (FMHM) with increasing strain\cite{m22}. The valley-dependent properties of monolayer $\mathrm{MoSi_2N_4}$, $\mathrm{WSi_2N_4}$ and $\mathrm{MoSi_2As_4}$ have been studied by the DFT calculations\cite{m23,m24}.

It's a natural idea to achieve Janus 2D materials in the new septuple-atomic-layer 2D $\mathrm{MA_2Z_4}$ family.
In this work, inspiring from the already synthesized  $\mathrm{MSi_2N_4}$ (M=Mo and W)  by introducing Si during CVD growth of $\mathrm{M_2N}$ (M=Mo and W)\cite{msn}, we construct the $\mathrm{MSiGeN_4}$ (M=Mo and W) monolayers, which may be achieved by  introducing Si and Ge during CVD growth of $\mathrm{M_2N}$ (M=Mo and W). Their  electronic structures, carrier mobilities and  piezoelectric properties  have been investigated, and show  distinct Rashba spin splitting and out-of-plane  piezoelectric polarizations  compared to $\mathrm{MSi_2N_4}$ (M=Mo and W) monolayers\cite{m21}.
It is found that the strain can effectively tune the electronic structures and piezoelectric properties of $\mathrm{MSiGeN_4}$ (M=Mo and W) monolayers.

The rest of the paper is organized as follows. In the next
section, we shall give our computational details and methods.
 In  the next few sections,  we shall present structural stabilities, electronic structures, carrier mobilities and piezoelectric properties of $\mathrm{MSiGeN_4}$ (M=Mo and W) monolayers, along with strain effects on their electronic structures and piezoelectric properties. Finally, we shall give our discussion and conclusions.

\begin{table}
\centering \caption{For $\mathrm{MSiGeN_4}$ (M=Mo and W) monolayers, the lattice constants $a_0$ ($\mathrm{{\AA}}$), the gaps with GGA and GGA+SOC (eV), and Rashba energy (meV). }\label{tab0}
  \begin{tabular*}{0.48\textwidth}{@{\extracolsep{\fill}}ccccc}
  \hline\hline
Name&$a_0$&  Gap& Gap-SOC&$E_R$\\\hline
$\mathrm{MoSiGeN_4}$&2.963&1.116&1.126&0.8 \\\hline
$\mathrm{WSiGeN_4}$&2.964&1.428&1.408&4.2 \\\hline\hline
\end{tabular*}
\end{table}

\section{Computational detail}
We perform DFT\cite{1} calculations for structural
relaxation and electronic structures by using
the Perdew-Burke-Ernzerhof generalized gradient approximation (PBE-GGA) for the exchange and correlation function,
as implemented in the Vienna ab initio simulation package
(VASP)\cite{pv1,pv2,pv3,pbe}. To describe the electron-ion interaction, we use  the projector augmented wave (PAW) method.
For energy band calculations of $\mathrm{MSiGeN_4}$ (M=Mo and W) monolayers, the SOC
is also taken into account.  A cutoff energy of 500 eV for the
plane wave basis set is used to ensure an
accurate DFT calculations. For the convergence
of  electronic self-consistent calculations, the total energy  convergence criterion is set
to $10^{-8}$ eV, and  the Hellmann-Feynman forces  on each atom are less than 0.0001 $\mathrm{eV.{\AA}^{-1}}$.
A vacuum spacing of more than 32 $\mathrm{{\AA}}$ is adopted to  decouple
the spurious interaction  between
the layers.

The coefficients of the elastic stiffness tensor  $C_{ij}$  and piezoelectric stress coefficients $e_{ij}$  are calculated by using strain-stress relationship (SSR) and   density functional perturbation theory (DFPT) method\cite{pv6}, respectively.
 The Brillouin zone sampling
is done using a Monkhorst-Pack mesh of 16$\times$16$\times$1  for $C_{ij}$, and  9$\times$16$\times$1 for $e_{ij}$.\
The 2D elastic coefficients $C^{2D}_{ij}$
 and   piezoelectric stress coefficients $e^{2D}_{ij}$
have been renormalized by the length of unit cell along z direction ($Lz$):  $C^{2D}_{ij}$=$Lz$$C^{3D}_{ij}$ and $e^{2D}_{ij}$=$Lz$$e^{3D}_{ij}$.
 The  phonon dispersion
spectrums are calculated by Phonopy code\cite{pv5} with a supercell
of 5$\times$5$\times$1 using the finite displacement method, and  a 3$\times$3$\times$1 k-mesh is employed. The kinetic energy cutoff is
set to 500 eV.  The constant energy contour plots of the spin
texture are calculated by the PYPROCAR code\cite{py}.

\begin{figure}
  \includegraphics[width=8cm]{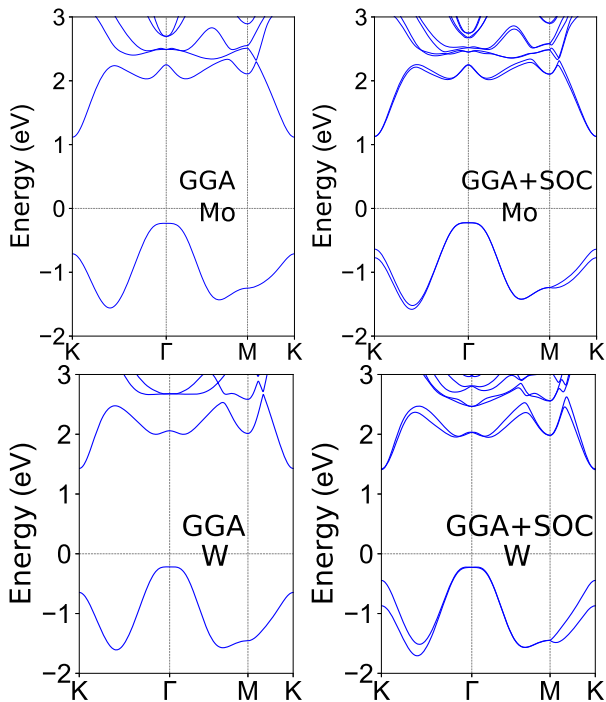}
\caption{(Color online) The energy band structures  of   $\mathrm{MSiGeN_4}$ (M=Mo and W)  using GGA  and GGA+SOC. }\label{t1-1}
\end{figure}

\begin{figure}
   \includegraphics[width=8.0cm]{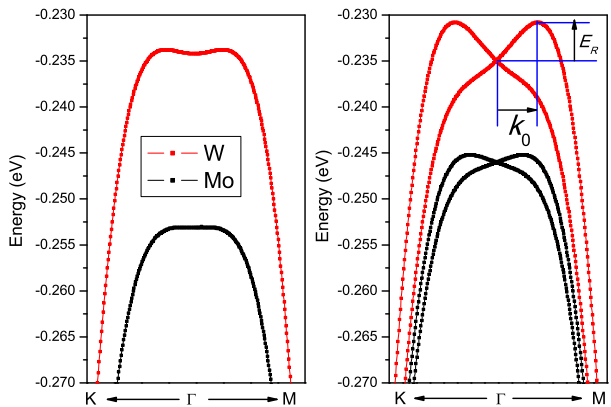}
  \caption{(Color online) The enlarged view of the valence bands  near the Fermi level for  $\mathrm{MSiGeN_4}$ (M=Mo and W) monolayers using GGA (Left)  and GGA+SOC (Right).}\label{t2-ra}
\end{figure}
\section{Structure and stability}
 The top view  and side view of  crystal structure of  $\mathrm{MSiGeN_4}$ (M=Mo and W) monolayers  are shown  in \autoref{t0}, and the rhombus primitive cell  and the rectangle supercell are shown. The structure of monolayer
 $\mathrm{MSiGeN_4}$ (M=Mo and W) could be regarded as a  $\mathrm{MN_2}$ layer  sandwiched by  Si-N and Ge-N
bilayers, which can be constructed by replacing the Si/Ge atoms of  top SiN/GeN bilayer in   $\mathrm{MSi_2N_4}$/ $\mathrm{MGe_2N_4}$ monolayer with Ge/N atoms.
If the  Si-N or Ge-N bilayers  is considered  as a whole, the $\mathrm{MSiGeN_4}$ (M=Mo and W) monolayers can be viewed as Janus 2D materials.
The symmetry of $\mathrm{MSiGeN_4}$ (M=Mo and W) monolayers (No.156) is lower than that of the $\mathrm{MSi_2N_4}$/ $\mathrm{MGe_2N_4}$  monolayer (No.187) due to
the lack of  the reflection symmetry with respect to the central  M atomic layer. The reduced symmetry can lead to many novel properties, such as Rashba spin splitting and out-of-plane  piezoelectric polarizations.

The optimized lattice constants of $\mathrm{MoSiGeN_4}$/$\mathrm{WSiGeN_4}$  is $a$=$b$=2.963/2.964 $\mathrm{{\AA}}$ with GGA, being between the ones of $\mathrm{MoSi_2N_4}$ (2.91  $\mathrm{{\AA}}$)/$\mathrm{WSi_2N_4}$ (2.91  $\mathrm{{\AA}}$) and $\mathrm{MoGe_2N_4}$ (3.02 $\mathrm{{\AA}}$)/ $\mathrm{WGe_2N_4}$(3.02 $\mathrm{{\AA}}$)\cite{msn,m20}.  The dynamical stability of the $\mathrm{MSiGeN_4}$ (M=Mo and W) monolayers  are tested
by analyzing the phonon spectra.  Their phonon band dispersions calculated along the high-symmetry
directions of the Brillouin zone are shown in \autoref{ph}.   The 18 optical and 3 acoustical phonon
branches as a total of 21 branches due to 7
atoms per cell are observed. It is clearly seen that the outlines of phonon band dispersions  between  $\mathrm{MoSiGeN_4}$ and  $\mathrm{WSiGeN_4}$ are very similar. It is noted that  the out-of-plane acoustic (ZA)  branch corresponding  to the out-of-plane vibrations deviates from linearity, which agrees well with  the conclusion that  the  ZA phonon branch should have quadratic dispersion, when the sheet is free of stress\cite{r1,r2}.
All phonon frequencies of the $\mathrm{MSiGeN_4}$ (M=Mo and W) monolayers are positive, which
confirms their dynamical stability,  and they  can exist as free-standing
2D materials.
\begin{figure*}
   \includegraphics[width=15.0cm]{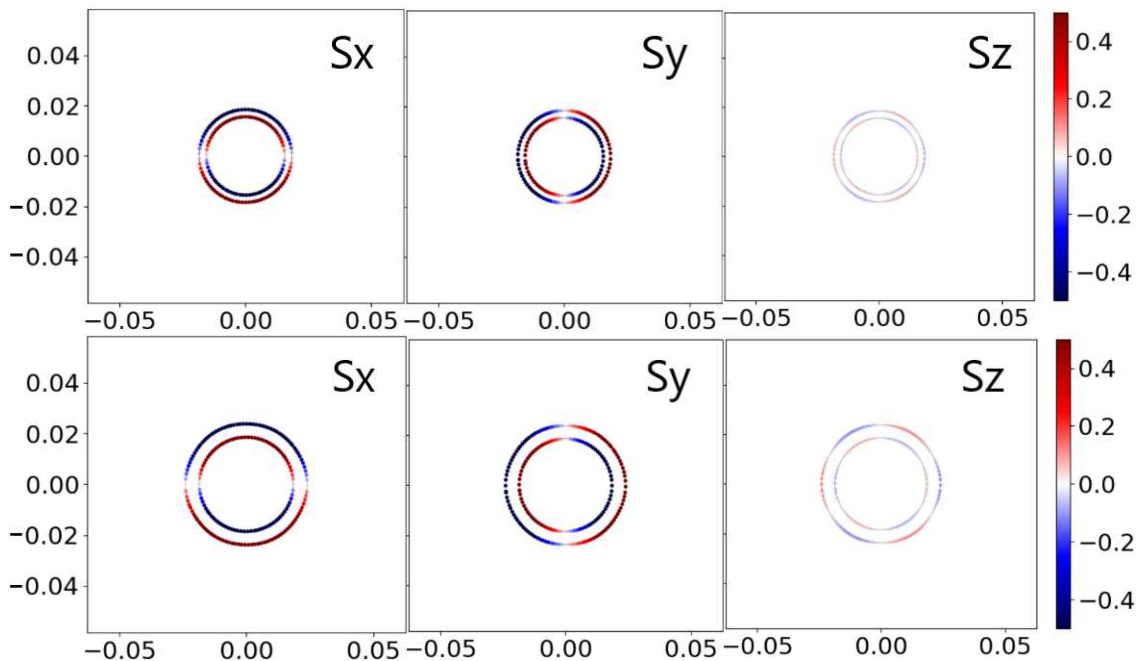}
  \caption{(Color online) Spin texture calculated in a $k_x-k_y$ plane centered at the $\Gamma$ point and at an
energy surface of -0.25 (-0.24) eV below the Fermi level for $\mathrm{MoSiGeN_4}$[Top] ($\mathrm{WSiGeN_4}$[Bottom]) monolayer. The red and blue colours show spin-up and spin-down states, respectively.}\label{t2}
\end{figure*}

It is important to check
the mechanical stability of the $\mathrm{MSiGeN_4}$ (M=Mo and W) monolayers by
elastic constants $C_{ij}$. The hexagonal symmetry leads to two  independent elastic
constants  $C_{11}$ and $C_{12}$ for $\mathrm{MSiGeN_4}$ (M=Mo and W) monolayers.
 The calculated   $C_{11}$=$C_{22}$=486.71 $\mathrm{Nm^{-1}}$/508.27 $\mathrm{Nm^{-1}}$ and $C_{12}$=144.14 $\mathrm{Nm^{-1}}$/147.21 $\mathrm{Nm^{-1}}$ for $\mathrm{MoSiGeN_4}$/$\mathrm{WSiGeN_4}$ monolayer.
 For hexagonal symmetry, the mechanical stability of a material should satisfy the  Born  criteria of mechanical stability\cite{ela}:
 \begin{equation}\label{e1}
C_{11}>0,~~ C_{66}>0
\end{equation}
where the $C_{66}$=($C_{11}$-$C_{12}$)/2.  The calculated $C_{ij}$ confirm  the mechanical stability of  $\mathrm{MSiGeN_4}$ (M=Mo and W) monolayers.
The Young's modulus $C_{2D}(\theta)$ can be calculated  on the basis of the elastic constants\cite{ela1}:
\begin{equation}\label{c2d}
C_{2D}(\theta)=\frac{C_{11}C_{22}-C_{12}^2}{C_{11}sin^4\theta+Asin^2\theta cos^2\theta+C_{22}cos^4\theta}
\end{equation}
where $A=(C_{11}C_{22}-C_{12}^2)/C_{66}-2C_{12}$.
 It is worth noting that $\mathrm{MSiGeN_4}$ (M=Mo and W) monolayers are  mechanically isotropic.
 The calculated $C_{2D}$ is 444.02 $\mathrm{Nm^{-1}}$/465.63 $\mathrm{Nm^{-1}}$ for  $\mathrm{MoSiGeN_4}$/$\mathrm{WSiGeN_4}$ monolayer, which are larger  than ones of most 2D materials\cite{ela2,ela3,ela4,ela5}, indicating that these monolayers are rigid.
 The Poisson's ratio $\nu(\theta)$ is also isotropic, and can be attained by:
 \begin{equation}\label{e1}
\nu^{2D}=\frac{C_{12}}{C_{11}}
\end{equation}
 The calculated  $\nu$ is 0.296/0.290 for  $\mathrm{MoSiGeN_4}$/$\mathrm{WSiGeN_4}$ monolayer.

 To verify the stability of the $\mathrm{MSiGeN_4}$ (M=Mo and W) monolayers at room temperature, ab initio
molecular dynamics (AIMD) simulations are carried out with a supercell
of size 4$\times$4$\times$1 for more than
3000 fs with a time step of 1 fs. The total energy fluctuations of $\mathrm{MSiGeN_4}$ (M=Mo and W) monolayers as a function of simulation time together with crystal structures at 300 K after the simulation for 3 ps are shown in FIG.1 of ESI.
  Calculated results show no obvious structural disruption with the total energy
 fluctuates being small after 3 ps at 300 K, which proves that $\mathrm{MSiGeN_4}$ (M=Mo and W) monolayers are thermodynamically
stable.

The dynamical, thermal  and mechanical  stability of the $\mathrm{MSiGeN_4}$ (M=Mo and W) monolayers are proved by
phonon calculations, AIMD and
elastic constants, suggesting the possible synthesis of
these monolayers. By introducing Si during CVD growth of $\mathrm{M_2N}$  (M=Mo and W), monolayer $\mathrm{MSi_2N_4}$ (M=Mo and W) have been
synthesized in experiment\cite{msn}. If the Si and Ge are simultaneously introduced during CVD growth of $\mathrm{M_2N}$  (M=Mo and W)  to passivate its surface, it is possible to achieve
$\mathrm{MSiGeN_4}$ (M=Mo and W) monolayers.

\begin{figure*}
  \includegraphics[width=12cm]{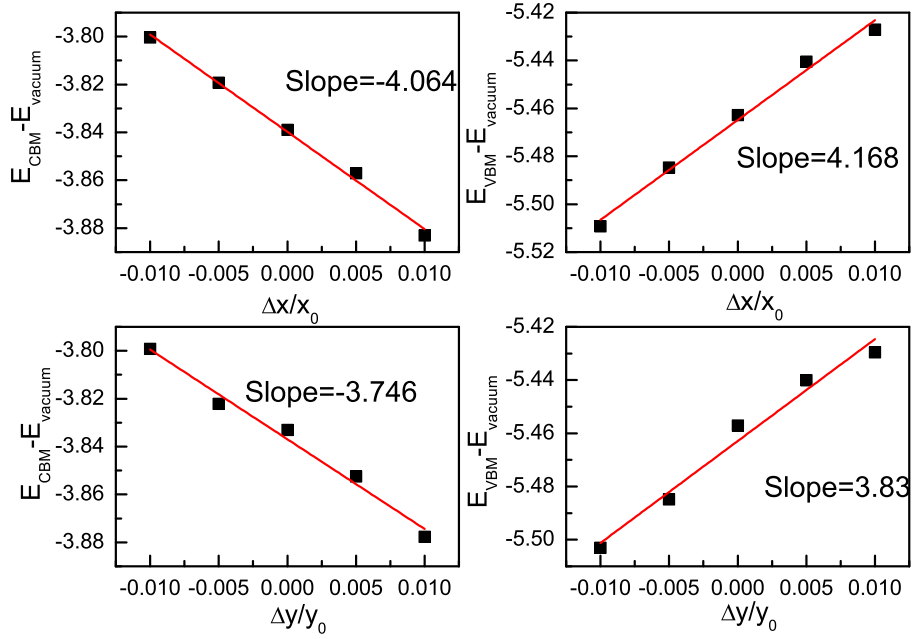}
  \caption{(Color online) The band energies of the VBM and CBM of $\mathrm{WSiGeN_4}$ monolayer with respect to the vacuum energy as a function of lattice dilation along both x and y directions using GGA+SOC. The red solid lines are linear fitting curves with fitted slopes as the DP constant.}\label{t4-dp}
\end{figure*}
\begin{figure*}
  \includegraphics[width=15cm]{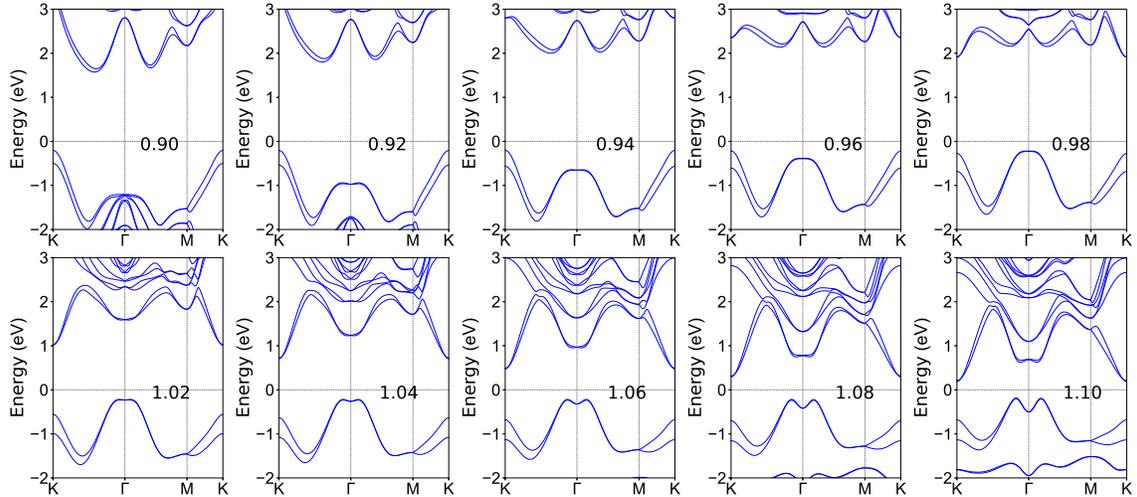}
\caption{(Color online) The energy band structures  of $\mathrm{WSiGeN_4}$ monolayer using GGA+SOC with $a/a_0$ changing from 0.90 to 1.10.}\label{t3}
\end{figure*}
\section{Electronic structure}
Due to containing transition metal in $\mathrm{MSiGeN_4}$ (M=Mo and W) monolayers, the SOC
is also taken into account. In fact, it has been proved that the SOC has  important effects on electronic structures of monolayer $\mathrm{MSi_2N_4}$ (M=Mo and W), which  exhibit rich spin-valley physics\cite{m20,m23,m24}. Therefore, the SOC is considered  for electronic structure  calculations of $\mathrm{MSiGeN_4}$ (M=Mo and W) monolayers, and  their energy band structures  with both GGA and GGA+SOC are plotted in \autoref{t1-1}.
Both GGA and GGA+SOC results show that $\mathrm{MSiGeN_4}$ (M=Mo and W) monolayers are indirect gap semiconductors with the  CBM at K point. To  accurately determine VBM, the enlarged views of the valence bands  near the Fermi level for  $\mathrm{MSiGeN_4}$ (M=Mo and W) monolayers using GGA   and GGA+SOC
are  plotted in \autoref{t2-ra}.  For GGA results, the valence bands  of  $\mathrm{MoSiGeN_4}$  around the $\Gamma$ point near the Fermi level are
 flat with the error less  than 1 meV, and the VBM  of  $\mathrm{WSiGeN_4}$ deviates slightly the $\Gamma$ point.
 Due to the intrinsic out-of-plane electric field induced by the mirror asymmetry, the  Rashba-type spin splitting around the $\Gamma$ point is observed, when the SOC is included. This gives rise to the deviation of VBM of $\mathrm{MSiGeN_4}$ (M=Mo and W) monolayers with GGA+SOC.
 It is found that the gap values of $\mathrm{MSiGeN_4}$ (M=Mo and W) monolayers between GGA and GGA+SOC are very close, and the related data are summarized in \autoref{tab0}.

From FIG.2 of ESI, the Zeeman-type spin
splitting around K/K1 point (the degenerate K and
K1 valleys ) in the valence bands near the Fermi level is observed  due to SOC together with inversion
symmetry breaking. The respective time-reversal symmetry requires that the spin splitting must be opposite at the two distinct valleys, which can be observed from FIG.2 of ESI. Moreover, due to the existence of the
horizontal mirror, they are fully spin-polarized in the out-of-plane direction (only $S_Z$ component), which is confirmed by our calculated results with FIG.2 of ESI being only $S_Z$ component.
$\mathrm{MSiGeN_4}$ (M=Mo and W) monolayers have conduction band valleys at K and K1.  Although the VBM  is not at the K/K1,  the valleys are still well defined and not far in energy. The similar results can be observed in monolayer $\mathrm{MoSi_2N_4}$, $\mathrm{WSi_2N_4}$ and $\mathrm{MoSi_2As_4}$\cite{m23,m24}.

The constant energy 2D contour plots
of spin texture calculated in a $k_x-k_y$ plane centered at the $\Gamma$ point are shown in \autoref{t2}.   The Rashba-type spin
splitting of spin-up (red) and spin-down (blue) electronic
bands can be distinctly observed.  The  2D Rashba spin splitting of valence bands gives rise to the concentric spin-texture circles with clockwise and counterclockwise rotating spin directions, respectively.
The concentric spin-texture circles are due to  the pure 2D Rashba spin splitting  in the valence bands.  It is found that  only in-plane $S_X$ and $S_Y$ spin components are
present in the Rashba spin split bands, without the presence of
any out-of-plane $S_Z$ component, which is also proved from FIG.2 of ESI.
The strength of the Rashba effect can be measured by three
key parameters: the Rashba energy ($E_R$), the Rashba momentum ($k_0$), and the Rashba constant ($\alpha_R$), and they can be connected by $\alpha_R$=2$E_R$/$k_0$. The $E_R$ and $k_0$ are shown in \autoref{t2-ra}.
We find that $E_R$, $k_0$ and $\alpha_R$  of  $\mathrm{MoSiGeN_4}$/$\mathrm{WSiGeN_4}$ monolayer are 0.8/4.2 meV, 0.048/0.076 $\mathrm{{\AA}}^{-1}$, and 0.033/0.111 eV$\mathrm{{\AA}}$.

\section{Carrier mobility and Piezoelectric properties}
 The carrier mobilities ($\mu_{2D}$) of  $\mathrm{MSiGeN_4}$ (M=Mo and W) monolayers are calculated by the deformation potential (DP) theory  proposed by Bardeen and Shockley\cite{dp}, which  is defined as:
\begin{equation}\label{u2d}
  \mu_{2D}=\frac{e\hbar^3C_{2D}}{K_BTm^*m_dE_l^2}
\end{equation}
where $T$,  $m^*$ is temperature and  the effective mass in the transport direction, and $m_d=\sqrt{m_xm_y}$ is the average effective mass.
The  elastic modulus $C_{2D}$ can be attained from $C_{ij}$.   In addition, $E_l$ is the DP constant defined by $E_l=\Delta E/\delta$
, where $\Delta E$ is the energy shift of the band edge of CBM or VBM with respect
to the vacuum level, and $\delta=\Delta l/l_0$ with applying uniaxial strain.

\begin{table*}
\centering \caption{For $\mathrm{MSiGeN_4}$ (M=Mo and W) monolayers, elastic modulus ($C_{2D}$) using GGA, effective mass ($m^*$) and deformation potential ($E_l$) using GGA+SOC, carrier mobility ($\mu_{2D}$)  at 300 K.}\label{tab3}
  \begin{tabular*}{0.96\textwidth}{@{\extracolsep{\fill}}ccccccc}
  \hline\hline
&Carrier type&    &$C_{2D}$ ($\mathrm{Nm^{-1}}$) & $m^*$ & $E_l$ (eV)& $\mu_{2D}$ ($\mathrm{cm^2V^{-1}s^{-1}}$)\\\hline\hline
&Electrons   & x&  444.02&        0.41&       -3.37         & 5205.14   \\
 $\mathrm{MoSiGeN_4}$& &y& 444.02&       0.38 &     -3.13          & 6573.25                                               \\
&Holes   & x&       444.02&      -10.66&     3.56        &6.56                                                 \\
        &     &y&   444.02&      -10.67&    3.39         &7.22                                  \\\hline\hline
&Electrons   & x&  465.63&        0.30&       -4.06        & 7046.80   \\
 $\mathrm{WSiGeN_4}$&&y&    465.63&       0.28 &      -3.75          & 8767.94                                               \\
&Holes   & x&       465.63&     -8.73&    4.17         &19.45                                                \\
         &    &y&   465.63&      -1.29&     3.83          &155.43                                 \\\hline\hline

\end{tabular*}
\end{table*}
\begin{figure}
   \includegraphics[width=7.0cm]{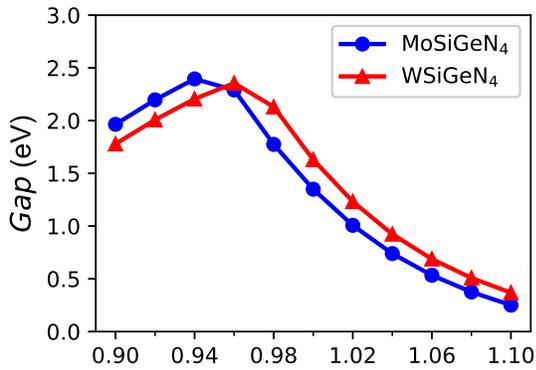}
  \caption{(Color online) The energy band gaps of  $\mathrm{MSiGeN_4}$ (M=Mo and W) monolayers as a function of  $a/a_0$ (0.90-1.10) by using GGA+SOC.}\label{t4}
\end{figure}

According to  DP theory,  we  calculate the carrier mobilities of $\mathrm{MSiGeN_4}$ (M=Mo and W) monolayers in both zigzag and armchair
directions  with armchair and zigzag being defined as x and y directions in \autoref{t0}.
The calculated effective masses for electrons  and holes of $\mathrm{MSiGeN_4}$ (M=Mo and W) monolayers with  GGA+SOC are shown in \autoref{tab3}. It is worth noting that the SOC has very important effects on the effective masses for holes due to different energy structures between GGA and GGA+SOC in \autoref{t2-ra}. The band energies of the VBM and CBM  with respect to the vacuum energy as a function of $\Delta x/x$ and $\Delta y/y$  are  plotted in \autoref{t4-dp} for $\mathrm{WSiGeN_4}$ monolayer, and FIG.3 of ESI for $\mathrm{MoSiGeN_4}$ monolayer. By linearly fitting these energy values, the DP constant $E_l$ can be attained.
The carrier mobilities of $\mathrm{MSiGeN_4}$ (M=Mo and W) monolayers for the electrons and holes  along x and y directions are attained on the basis of the calculated $m^*$, $C_{2D}$ and $E_l$. The related data are summarized in \autoref{tab3}.
The very strong anisotropy   of predicted carrier mobilities between  electrons and holes is observed, and the electron  carrier mobilities of $\mathrm{MSiGeN_4}$ (M=Mo and W)  monolayer are very  higher than those of holes. The electron  carrier mobilities of $\mathrm{MoSiGeN_4}$ ($\mathrm{WSiGeN_4}$) along x and y directions are up to 5205 $\mathrm{cm^2V^{-1}s^{-1}}$  (7047 $\mathrm{cm^2V^{-1}s^{-1}}$) and 6573 $\mathrm{cm^2V^{-1}s^{-1}}$ (8768 $\mathrm{cm^2V^{-1}s^{-1}}$).

Next, we investigate the  piezoelectric properties of  $\mathrm{MSiGeN_4}$ (M=Mo and W) monolayers.
Performing symmetry analysis, due to a $3m$ point-group symmetry,  the piezoelectric stress   and strain tensors, and elastic   tensor
can be  reduced into\cite{ela2}:
 \begin{equation}\label{pe1-1}
 e=\left(
    \begin{array}{ccc}
      e_{11} & -e_{11} & 0 \\
     0 & 0 & -e_{11} \\
      e_{31} & e_{31} & 0 \\
    \end{array}
  \right)
    \end{equation}

  \begin{equation}\label{pe1-2}
  d= \left(
    \begin{array}{ccc}
      d_{11} & -d_{11} & 0 \\
      0 & 0 & -2d_{11} \\
      d_{31} & d_{31} &0 \\
    \end{array}
  \right)
\end{equation}

 \begin{equation}\label{pe1-3}
   C=\left(
    \begin{array}{ccc}
      C_{11} & C_{12} & 0 \\
     C_{12} & C_{11} &0 \\
      0 & 0 & (C_{11}-C_{12})/2 \\
    \end{array}
  \right)
\end{equation}
Here, the independent $d_{11}$ and $d_{31}$ are derived by $e_{ik}=d_{ij}C_{jk}$:
\begin{equation}\label{pe2}
    d_{11}=\frac{e_{11}}{C_{11}-C_{12}}~~~and~~~d_{31}=\frac{e_{31}}{C_{11}+C_{12}}
\end{equation}

\begin{table}
\centering \caption{Piezoelectric coefficients $e_{11}(d_{11})$ and $e_{31} (d_{31})$ of $\mathrm{MSiGeN_4}$, $\mathrm{MSi_2N_4}$ and  $\mathrm{MGe_2N_4}$ (M=Mo and W) monolayers, and the unit is $10^{-10}$C/m (pm/V). }\label{tab5}
  \begin{tabular*}{0.48\textwidth}{@{\extracolsep{\fill}}ccccc}
  \hline\hline
Name & $e_{11}$ & $d_{11}$& $e_{31}$&$d_{31}$\\\hline
 $\mathrm{MoSi_2N_4}$& 4.395&1.144 &- & -  \\\hline
$\mathrm{MoSiGeN_4}$&5.116& 1.494&-0.087&-0.014                               \\\hline
 $\mathrm{MoGe_2N_4}$&5.621& 1.846&- & -\\\hline\hline
$\mathrm{WSi_2N_4}$& 3.138&      0.778   &- & - \\\hline
 $\mathrm{WSiGeN_4}$&3.790& 1.050&0.073 & 0.011                                \\\hline
 $\mathrm{WGe_2N_4}$&4.218& 1.306&- & -\\\hline\hline
\end{tabular*}
\end{table}

For $e_{ij}$, the  orthorhombic  supercell of  $\mathrm{MSiGeN_4}$ (M=Mo and W) monolayers (in \autoref{t0}) is adopted, and the calculated $e_{ij}$ and $d_{ij}$  are summarized in \autoref{tab5}, along with ones of $\mathrm{MA_2N_4}$ (M=Mo and W; A=Si and Ge) monolayer.
With respect to the central M
atomic plane, the $\mathrm{MA_2N_4}$ (M=Mo and W; A=Si and Ge) monolayer
possess a reflection symmetry  due to $D_{3h}$ symmetry, which leads to that they have only  in-plane piezoelectricity.
For $\mathrm{MSiGeN_4}$ (M=Mo and W) monolayers,  the difference in
atomic sizes and electronegativities of the second and sixth layer atoms  breaks the reflection symmetry along the
vertical direction, giving rise to  a low degree of $3m$ symmetry.
Therefore, both in-plane  and
vertical piezoelectricity are allowed in  $\mathrm{MSiGeN_4}$ (M=Mo and W) monolayers, when they are subject to a uniaxial in-plane strain.
It is clearly seen that both $e_{11}$ and $d_{11}$ increase with increasing atomic mass from  $\mathrm{MSi_2N_4}$ (M=Mo and W) to  $\mathrm{MSiGeN_4}$ (M=Mo and W) to  $\mathrm{MGe_2N_4}$ (M=Mo and W). It is found that  the $\mathrm{MoSiGeN_4}$ and $\mathrm{MoA_2N_4}$ (A=Si and Ge)
 monolayers have higher $e_{11}$/$d_{11}$ values than $\mathrm{WSiGeN_4}$ and $\mathrm{WA_2N_4}$ (A=Si and Ge) monolayers.
 For a given metal element M, the
monolayers containing heavier column IV element have larger
$e_{11}$/$d_{11}$ values. More significantly, the  $\mathrm{MSiGeN_4}$ (M=Mo and W) monolayers possess the
vertical piezoelectric effect, which can be described by $e_{31}$/$d_{31}$. However, they are smaller by two orders of magnitude compared to $e_{11}$/$d_{11}$. Similar phenomenon can be observed in  Janus MXY (M =
Mo or W, X/Y = S, Se, or Te) monolayer\cite{ela2}.

\begin{figure}
   \includegraphics[width=7.0cm]{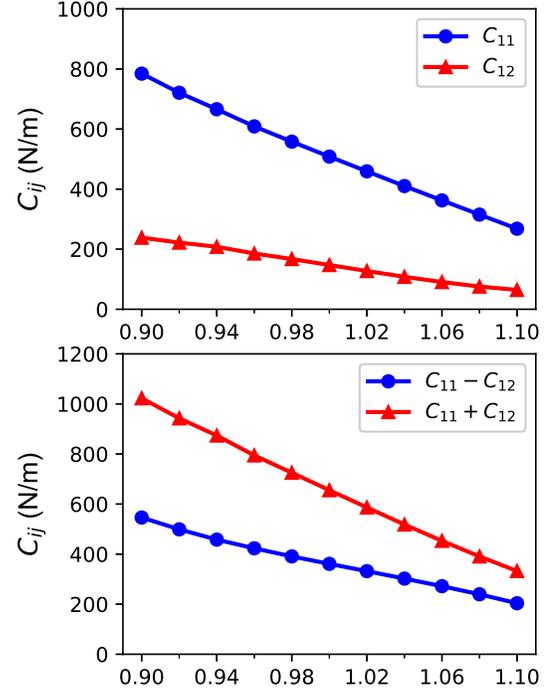}\\
  \caption{(Color online) For monolayer   $\mathrm{WSiGeN_4}$,  the elastic constants  $C_{ij}$ with the application of  biaxial strain (0.90 to 1.10).}\label{t5}
\end{figure}

\begin{figure}
   \includegraphics[width=7.0cm]{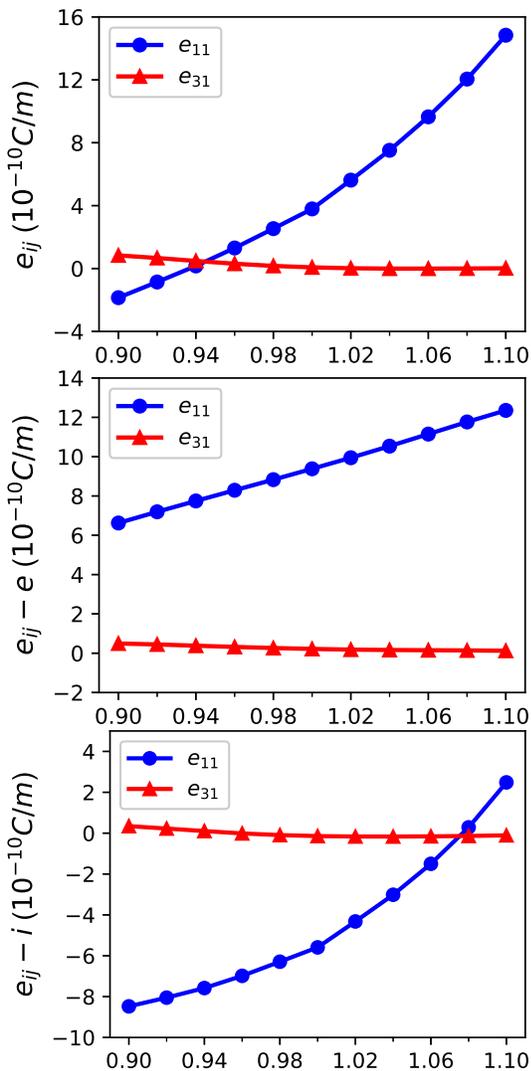}
  \caption{(Color online) For monolayer $\mathrm{WSiGeN_4}$, the piezoelectric stress coefficients  $e_{11}$ and  $e_{31}$ along with  the ionic contribution and electronic contribution to  $e_{11}$ and  $e_{31}$  with the application of  biaxial strain (0.90 to 1.10).}\label{t6}
\end{figure}

\begin{figure}
   \includegraphics[width=7.0cm]{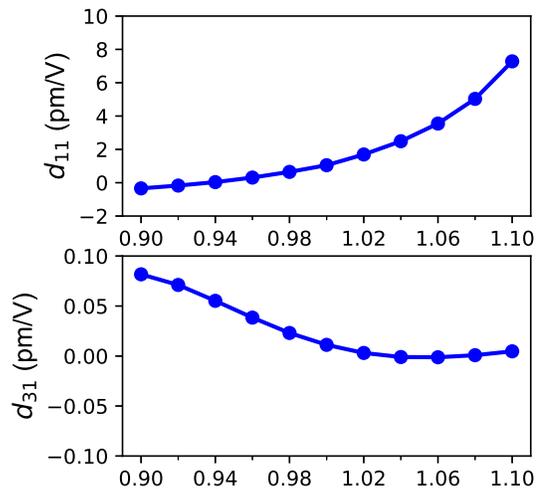}
  \caption{(Color online) For monolayer  $\mathrm{WSiGeN_4}$, the piezoelectric strain coefficients  $d_{11}$ and $d_{31}$ with the application of  biaxial strain (0.90 to 1.10).}\label{t7}
\end{figure}

\section{Strain effects}
 It has been proved that  the electronic structures, topological properties, transport and   piezoelectric  properties  of  2D materials  can be  effectively tuned by strain\cite{m12,m13,m14,m15,r6,r7,r8}.
 Here,  we use $a/a_0$ to examine the effects of  biaxial strain on the electronic structures of   $\mathrm{MSiGeN_4}$ (M=Mo and W) monolayers,  where $a$ and $a_0$ are the strained and  unstrained lattice constant with $a/a_0$$<$1 ($a/a_0$$>$1) being compressive (tensile) strain,
 The energy band structures of  $\mathrm{WSiGeN_4}$ with $a/a_0$ from 0.90 to 1.10 are  plotted  in \autoref{t3}, and the related energy band structures are shown in FIG.4 of ESI for $\mathrm{MoSiGeN_4}$. The energy band gaps of $\mathrm{MSiGeN_4}$ (M=Mo and W) monolayers as a function of $a/a_0$   are shown in \autoref{t4}.  It is found that  the energy band gap of $\mathrm{MSiGeN_4}$ (M=Mo and W) monolayers firstly  increases with increasing $a/a_0$,  and then decreases.  The up-and down trend of gap  can also be observed in many 2D materials, like  Janus TMD monolayers\cite{q5} and GeS\cite{q5-11}.
 The compressive strain can make  conduction band extrema (CBE) of $\mathrm{WSiGeN_4}$  monolayer converge, especially for 0.96 and 0.98 strains.
 The conduction bands  convergence is in favour of n-type Seebeck coefficient.  The compressive strain can make K point become  VBM, which is very useful for manipulating valley pseudospin.  The compressive strain produces another effect that the CBM changes from K point to one point along K-$\Gamma$ line. Similar strain effects on electronic structures of $\mathrm{MoSiGeN_4}$ can be found. It is noted that  $\mathrm{MSiGeN_4}$ (M=Mo and W) monolayers in considered strain range are all semiconductors, which is useful for their  piezoelectric application  with strain.

  The piezoelectric strain coefficients of   $\mathrm{MSiGeN_4}$ (M=Mo and W) monolayers are very small,  and strain engineering may be an effective way  to enhance their piezoelectric properties. Next, we consider the strain effects on piezoelectric properties of  $\mathrm{MSiGeN_4}$ (M=Mo and W) monolayers. The elastic constants ($C_{11}$, $C_{12}$,  $C_{11}$-$C_{12}$ and  $C_{11}$+$C_{12}$),  piezoelectric  stress  coefficients  ($e_{11}$ and $e_{31}$ along  the ionic  and electronic contributions), and  piezoelectric  strain  coefficients ($d_{11}$ and $d_{31}$) of monolayer $\mathrm{WSiGeN_4}$ as a function of  biaxial  strain are plotted in \autoref{t5}, \autoref{t6} and \autoref{t7}, respectively.
 For $\mathrm{MoSiGeN_4}$, these are shown in FIG.5, FIG.6 and FIG.7 of ESI, respectively.
 With strain from 0.90 to 1.10, the $d_{11}$  increases due to decreased $C_{11}$-$C_{12}$  and enhanced $e_{11}$ based on \autoref{pe2}.
 At 10\% strain, the $d_{11}$ of   $\mathrm{WSiGeN_4}$ ($\mathrm{MoSiGeN_4}$) is  7.282 pm/V (8.081 pm/V), which is about seven times (five times) as large as  unstrained one  of 1.050 pm/V (1.494 pm/V). It is found that both ionic  and electronic parts have positive contribution to $e_{11}$ with increasing tensile strain.  Similar  biaxial  strain-enhanced  $d_{11}$ can be observed  in monolayer $\mathrm{MoSi_2N_4}$,  g-$\mathrm{C_3N_4}$ and $\mathrm{MoS_2}$ \cite{m21,gsd}. It is observed that the compressive strain can improve the $d_{31}$ (absolute value) of $\mathrm{MSiGeN_4}$ (M=Mo and W) monolayers due to enhanced $e_{31}$ (absolute value), and the $d_{31}$ can be improved to 0.082 pm/V (-0.086 pm/V) for  $\mathrm{WSiGeN_4}$ ($\mathrm{MoSiGeN_4}$) at 0.90 strain. Finally, it is found that   the$\mathrm{MSiGeN_4}$ (M=Mo and W) monolayers are mechanically stable in the considered strain range, based on calculated elastic constants satisfying  the mechanical stability criteria.

\section{Discussions and Conclusion}
The  $\mathrm{MSi_2N_4}$ (M = Mo, W) monolayers have been recently synthesized, which are  grown by passivating the surface dangling bonds of  $\mathrm{MN_2}$ (M = Mo, W) layer with Si-N tetrahedra when introducing elemental Si\cite{msn}.
Thus, it is possible to achieve Janus $\mathrm{MSiGeN_4}$ (M=Mo and W) monolayers by  simultaneously introducing Si and Ge elements during CVD growth of nonlayered $\mathrm{MN_2}$ (M = Mo, W) to passivate its surface.  Compared to $\mathrm{MSi_2N_4}$ (M = Mo, W) monolayers, the most important difference is that  Janus $\mathrm{MSiGeN_4}$ (M=Mo and W) monolayers have out-of-plane piezoelectric polarization and  Rashba effect due to their out-of-plane
asymmetry. Although their out-of-plane piezoelectric polarization and  Rashba effect are very weak, our works open a new avenue to achieve Janus materials in the new 2D  $\mathrm{MA_2Z_4}$ family.

 In summary, we investigate the electronic structures, carrier mobilities, piezoelectric properties  of $\mathrm{MSiGeN_4}$ (M=Mo and W) monolayers by  the reliable first-principle calculations. They are  found to exhibit mechanical,  thermodynamical  and dynamic stability, and high experimental feasibility.
 It is found that $\mathrm{MSiGeN_4}$ (M=Mo and W) monolayers are  indirect gap semiconductors. When the SOC is considered, the  Rashba effect can be observed in the valence bands  of   $\mathrm{MSiGeN_4}$ (M=Mo and W) monolayers. Their electron   mobilities are very high due to very light electron effective masses.  The $e_{11}/d_{11}$ of $\mathrm{MSiGeN_4}$ (M=Mo and W) monolayers can be  induced by a
uniaxial strain in the basal plane, similar to $\mathrm{MSi_2N_4}$ (M = Mo, W) monolayers.  In addition to this,  a vertical piezoelectric polarization $e_{31}/d_{31}$  can be produced upon application of uniaxial or biaxial strains due to the lack of reflection symmetry
with respect to  M atomic layer. Calculated results show that compressive strain can change  the positions of CBM and VBM of $\mathrm{MSiGeN_4}$ (M=Mo and W) monolayers, and tune  the strength of conduction bands convergence.  It is also found that  biaxial  strain can enhance
 $d_{11}$ [$d_{31}$ (absolute values)] of  $\mathrm{MSiGeN_4}$ (M=Mo and W) monolayers by tensile [compressive] strain.
 Our works will  stimulate further experimental studies to achieve $\mathrm{MSiGeN_4}$ (M=Mo and W) monolayers, and  will motivate farther exploration on Janus monolayers in new 2D $\mathrm{MA_2Z_4}$ family.
\\
\\
\textbf{Conflicts of interest}
\\
There are no conflicts to declare.

\begin{acknowledgments}
This work is supported by the Natural Science Foundation of Shaanxi Provincial Department of Education (19JK0809). We are grateful to the Advanced Analysis and Computation Center of China University of Mining and Technology (CUMT) for the award of CPU hours and WIEN2k/VASP software to accomplish this work.
\end{acknowledgments}

\end{document}